\documentclass[10pt,twocolumn]{article}
\usepackage{graphicx} 
\usepackage{dblfloatfix}
\usepackage{amsmath,amssymb} 
\usepackage{bm} 
\usepackage{wrapfig} 
\usepackage{url}

\newcommand{\JournalTitle}[1]{#1}

\begin{document} 

\title{Multiple self-similar crossovers in a hydrodynamic interface} 

\author{ Wanjing Gu and Ko Okumura\\ \small Department of Physics and Soft Matter Center, Ochanomizu University,\\ \small 2-1-1 Ohtsuka, Bunkyo-ku, Tokyo 112-8610, Japan\\ \small E-mail: okumura@phys.ocha.ac.jp } 

\date{\today} 

\maketitle

\begin{abstract}

Self-similar dynamics are often described in terms of a single asymptotic
scaling regime. Here we experimentally investigate the post-breakup recovery
of a hydrodynamic interface over substantially wider temporal and spatial
ranges than previously accessible. Beyond the previously identified scaling
behavior, we find four successive self-similar regimes characterized by the $%
R_{\beta }${}-scaling with $\beta =1/6$, 1/2, 1, and 2, connected by three
distinct crossovers. All four regimes exhibit robust collapse near the
interface tip, while geometrical dependence becomes detectable farther from
the singular region. This systematic variation across space and scaling
regimes is consistent with scale separation and enhanced universality closer
to the breakup singularity. The characteristic scales associated with the
different $R_{\beta }${}{}-scalings become comparable at the common
geometric scale $z_{m}\approx R$. These findings demonstrate that
self-similar dynamics can undergo multiple successive crossovers among
distinct scaling regimes rather than directly approaching a single
asymptotic self-similar state.

\end{abstract}


\vspace{1em}



Universality is one of the central organizing principles of
modern science. Systems that differ substantially in their microscopic
details or experimental conditions can nevertheless exhibit common scaling
behavior described by universal laws and functions. Such behavior appears
across critical phenomena, nonlinear diffusion, stochastic growth, singular
fluid dynamics, biological pattern formation, and many other nonlinear
systems. In many cases, universality manifests itself through self-similar
dynamics, where evolving profiles collapse onto a single master curve after
appropriate rescaling \cite{Barenblatt, eggers2015singularities, Goldenfeld,
Cardy, barabasi1995fractal, cross1993pattern, murray2002mathematical1,
murray2002mathematical2}.

A common assumption underlying most studies of self-similarity is that the
dynamics ultimately approach a single asymptotic scaling regime
characterized by fixed scaling exponents and a unique scaling function \cite%
{Barenblatt, eggers2015singularities, Goldenfeld, Cardy, barabasi1995fractal}%
. This perspective has been remarkably successful in describing
singularities, coarsening processes, growth phenomena, and intermediate
asymptotics throughout physics \cite{Barenblatt, eggers2015singularities,
Goldenfeld, Cardy, barabasi1995fractal}.

While crossover phenomena are widely recognized and play important roles in
many areas of physics, including critical phenomena \cite{Goldenfeld, Cardy}%
, they are typically viewed as transitions between a small number of
distinct scaling regimes. For example, transitions between distinct
self-similar regimes have been observed in fluid-interface singularities,
including an oscillatory transition during liquid pinch-off \cite%
{lagarde2018oscillating} and a crossover between two self-similar regimes in
confined bubble pinch-off \cite{pahlavan2019restoring}. However, much less
is known about whether a single dynamical process can exhibit multiple
successive crossovers among distinct self-similar regimes.

Classical studies of fluid-interface singularities demonstrated that
asymptotic dynamics can lose memory of initial conditions and system
parameters and converge toward universal self-similar states \cite%
{1993PRLEggersPinchoff, 1994ScienceNagelDropFallingFaucet}. More recent work
showed that memory of initial conditions and system parameters may persist
in certain situations, leading to departures from universality \cite%
{2003ScienceNagelMemoryDropBreakup, pahlavan2019restoring}. Motivated by
these observations, Yoshino and Okumura recently proposed the concept of 
\textit{incomplete universality}, in which asymptotic self-similar dynamics
preserve partial information about geometrical confinement \cite%
{yoshino2025partial}. These developments raise a broader question: can
self-similar dynamics proceed through multiple successive crossovers among
distinct scaling regimes?

Recovery dynamics following topological transitions provide a particularly
suitable setting for addressing this question. Self-similar dynamics
associated with topological changes have been extensively studied in fluid
interfaces, including viscous-drop coalescence, substrate-mediated
coalescence, electrically driven coalescence, and thin-film coalescence \cite%
{BirdRistenpartBelmonteStone2009, YokotaPNAS2011, hernandez2012symmetric,
kaneelil2022three}. Unlike singular collapse processes, recovery dynamics
can be tracked continuously over extended temporal and spatial domains.
Consequently, they offer a unique opportunity to investigate how
self-similar states emerge and evolve during relaxation. Previous
experiments on air entrained into viscous liquids in quasi-two-dimensional
geometries revealed universal recovery dynamics characterized by distinct
self-similar scalings \cite{yoshino2025partial, JPS2023autumn}. However, the
accessible observation window was insufficient to determine whether
crossover structures existed.

\begin{figure*}[!t]
\centering\includegraphics[width=0.7\textwidth]{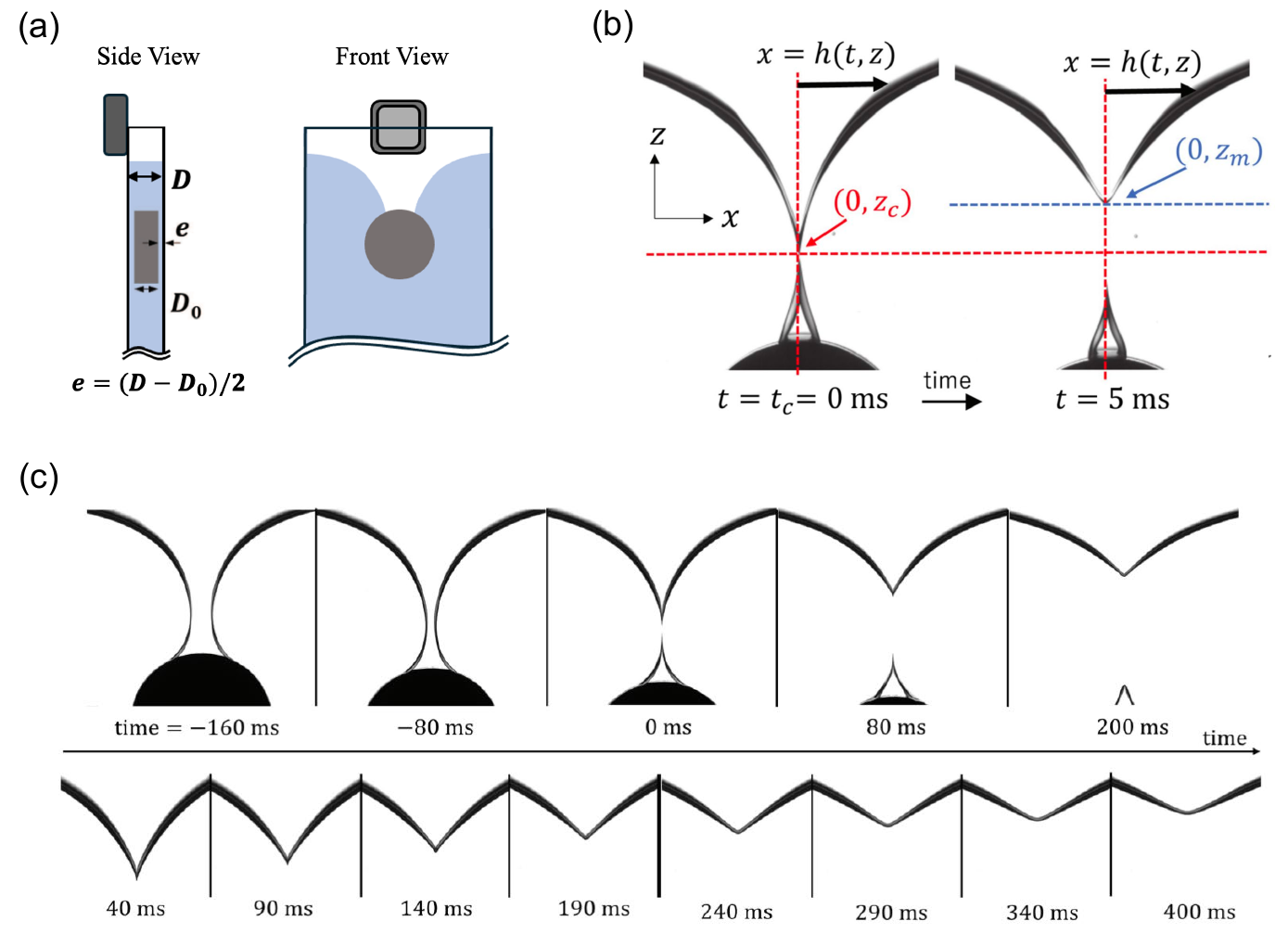}
\caption{(a) Experimental set-up with an electromagnet: Side and front views
of the falling disk inside the Hele-Shaw cell. (b) Coordinate system adopted
for the description of interface profiles after breakup. (c) Representative
snapshots of the interface evolution before and after breakup. The
experimental parameters are given by $(e,D_{0},R,\protect\nu )=(0.5$ mm$,3\,%
\text{mm},10\,\text{mm},10\,\text{St})$. Items (b) and (c) are adapted from
Yoshino \& Okumura, Phys. Rev. Research 2025 (CC BY 4.0).}
\label{Fig1}
\end{figure*}

Here we extend these experiments over substantially broader temporal and
spatial ranges and uncover multiple self-similar crossovers during interface
recovery following breakup. Rather than approaching a single asymptotic
self-similar state through a single dominant transition, the interface
evolves successively through four distinct scaling regimes. Across all four
regimes, the profiles exhibit robust universality near the interface tip,
while geometrical dependence becomes detectable farther from the singular
region. The systematic variation of this dependence across space and scaling
regimes suggests roles for both scale separation and the distance from the
breakup singularity. These observations reveal that a single recovery
process can undergo multiple successive crossovers among distinct
self-similar regimes while retaining enhanced universality near the singular
region. Remarkably, self-similarity itself persists through these
crossovers, while the scaling form governing the self-similar dynamics
changes successively. More broadly, these findings suggest that self-similar
dynamics can possess a richer crossover structure than is captured by the
conventional picture of direct evolution toward a single asymptotic scaling
state.

\section{Experimental}

As shown in Fig.~\ref{Fig1}(a), a Hele-Shaw cell (width 10 cm, height 15 cm,
and thickness $D=1.5$ to 4 mm) is filled with viscous oil
(polydimethylsiloxane, PDMS), whose kinematic viscosity ranges from $1$ to $%
50$ St. A stainless-steel disk (SUS430) is then released into the cell from
the horizontal air-liquid interface, entraining a sheet of air into the oil
phase. The entrained air subsequently undergoes topological breakup, after
which the interface gradually recovers toward a horizontal configuration.
The post-breakup recovery dynamics forms the focus of the present study.

The disk thickness $D_{0}$ ranged from $1$ to $3$ mm, while the disk radius $%
R$ varied between $10$ and $12.5$ mm. When the disk is placed in the
Hele-Shaw cell, a thin liquid film remains between the disk and the walls.
Its thickness is estimated as

\begin{equation}
e=(D-D_{0})/2  \label{eq:e}
\end{equation}%
The parameter $e$ plays an important role in previous observations of
incomplete universality and is systematically varied in the present study.

Compared with the previous study, experimental reproducibility was
substantially improved through an automated release system (Appendix \ref%
{App1}). Interface evolution was recorded with a high-speed camera at
1000--2000 frames per second (fps). The breakup time $t=t_{c}$ was
determined from two consecutive frames bracketing pinch-off, yielding a
temporal uncertainty below 1 ms. Experiments were performed under the
parameter sets listed in Table \ref{tab:exp_conditions}. 
\begin{table}[h]
\caption{Summary of the experimental conditions investigated in this study.
The length scales, $e$, $D_{0}$, and $R$, are given in mm, and the kinematic
viscosity $\protect\nu $ is given in St, throughout this article.}
\label{tab:exp_conditions}\centering
\par
\begin{tabular}{cl}
\hline\hline
Index & \multicolumn{1}{c}{$(e,D_0,R,\nu)$} \\ \hline
1 & $(0.50,3.0,12.5,10)$ \\ 
2 & $(0.50,2.0,10.0,10)$ \\ 
3 & $(0.75,2.5,10.0,10)$ \\ 
4 & $(1.00,2.0,10.0,10)$ \\ 
5 & $(0.50,2.5,10.0,10)$ \\ \hline\hline
\end{tabular}%
\end{table}

\begin{figure*}[!t]
\centering\includegraphics[width=0.8\textwidth]{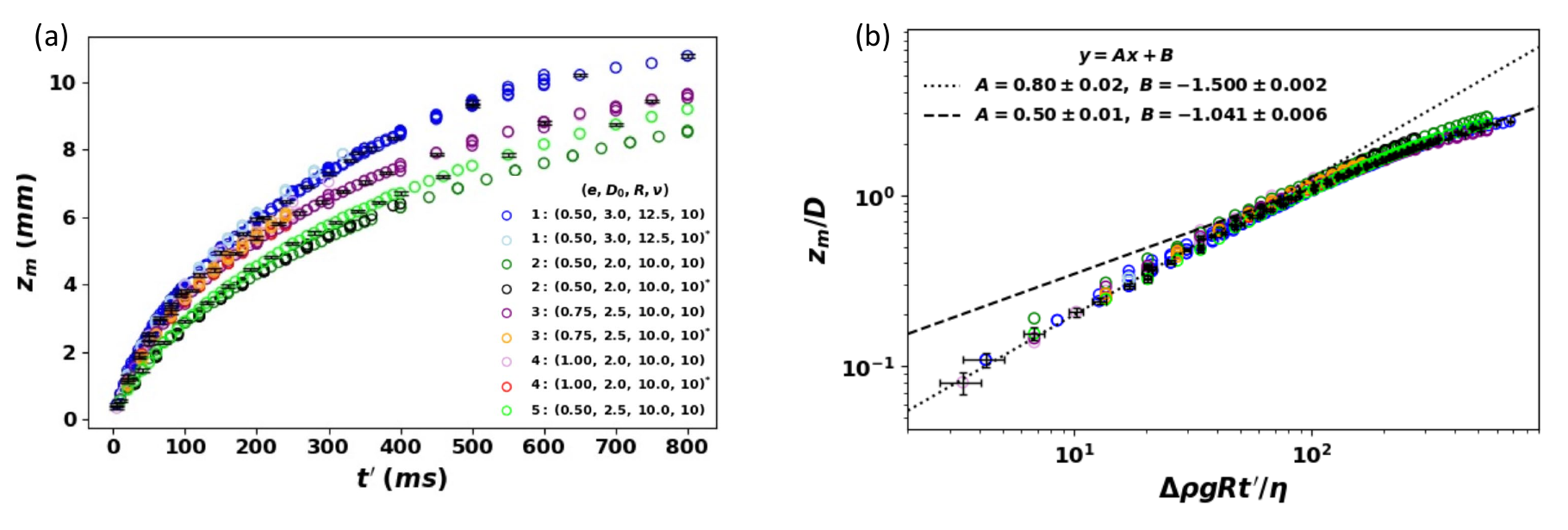}
\caption{(a) $z_{m}$ vs $t^{\prime }$ for multiple parameter sets. (b)
Dimensionless log-log plot of (a). Dotted and dashed lines in (b) represent
power-law fits with slopes of $0.80\pm 0.02$ and $0.50\pm 0.01$. Parameter
sets labeled as $(e,D_{0},R,\protect\nu )$ in the legends correspond to
those in Table~\protect\ref{tab:exp_conditions}. Representative error bars
for datasets 1 and 2 are shown in (a) and (b), corresponding to the spatial
and temporal resolutions ($\pm 1$ pixel and $\pm 1$ ms).}
\label{scaling_verify}
\end{figure*}

\section{Results}

We use the coordinate system illustrated in Fig.~\ref{Fig1}(b) throughout
this article. Representative interface profiles during recovery are shown in
Fig.~\ref{Fig1}(c). Immediately after breakup, the interface exhibits a
sharp cusp-like tip. As the recovery proceeds, the tip gradually rounds and
undergoes a sequence of morphological changes. By extending the observation
window beyond that of the previous study \cite{yoshino2025partial}, we
uncover multiple self-similar crossovers during interface recovery.

\subsection{Characteristic Length Scale $z_{m}(t)$}

\label{zm_scaling}

As a starting point, we revisit the scaling behavior of the characteristic
tip position $z_{m}(t)$ defined in Fig.~\ref{Fig1}(b), which provides the
fundamental length scale for all subsequent analyses. For $z_{m}(t)$, the
following scaling structure has been reported in a previous study by Yoshino
and Okumura \cite{yoshino2025partial}:

\begin{equation}
\frac{z_{m}(t)}{D}=f\left( \frac{\Delta \rho gRt^{\prime }}{\eta }\right)
\label{eq:yoshino_scaling_after}
\end{equation}%
where the elapsed time after breakup is defined as $t^{\prime }=t-t_{c}$
with $t_{c}$ denoting the breakup time. $\Delta \rho $ is the density
difference between the metal disk and viscous fluid. Furthermore, they found
a crossover between two scaling regimes: 
\begin{equation}
\frac{z_{m}}{D}\sim \left( \frac{\Delta \rho gRt^{\prime }}{\eta }\right)
^{\alpha }  \label{eq3}
\end{equation}%
where the exponent was found to be $\alpha \simeq 4/5$ and 1/2 in the
early-time and late-time regimes, respectively.

Fig.~\ref{scaling_verify}(a) presents the temporal evolution of the
characteristic length $z_{m}$ obtained in the present study for all
parameter sets specified in Table~\ref{tab:exp_conditions} and compares them
with the corresponding data reported by Yoshino and Okumura \cite%
{yoshino2025partial}, which are shown by the symbols marked with * in the
legend. In the plot, no meaningful difference is observed between the data
obtained with the new electromagnetic release system in the present study
and those with manual release in the previous study \cite{yoshino2025partial}%
. Correspondingly, the log-log plot in Fig.~\ref{scaling_verify}(b) confirms
the scaling structure in Eq.~(\ref{eq:yoshino_scaling_after}) and the
scaling crossover from $\alpha \approx 4/5$ to 1/2, established in the
previous study \cite{yoshino2025partial}. The possible influence of the
electromagnetic release system was further examined by varying the applied
voltage from 16 to 24 V. No measurable dependence of the post-breakup
dynamics on the applied voltage was observed, confirming that the release
procedure does not measurably affect the recovery dynamics.

Using all the data shown in Fig.~\ref{scaling_verify}, we obtained $\alpha
=0.80\pm 0.02$ in the first regime and $\alpha =0.50\pm 0.01$ in the second
by fitting $(x,y)$ $=$ $(\log _{10}(\Delta \rho gRt^{\prime }/\eta )$, $\log
_{10}(z_{m}/D))$ to $y=Ax+B$ over the ranges $x\in \lbrack 3,70]$ and $%
[70,800]$, respectively, indicating a crossover around $x\approx 70$. Here, $%
A$ corresponds to the exponent $\alpha $, while $B$, obtained as $-1.500\pm
0.002$ in the first regime and $-1.041\pm 0.006$ in the second, determines
the coefficient for Eq.~(\ref{eq3}).

Importantly, the characteristic length $z_{m}$ derived from the previously
known scaling law provides the basis for uncovering the additional
self-similar structures reported below.

\begin{figure*}[!t]
\centering\includegraphics[width=\textwidth]{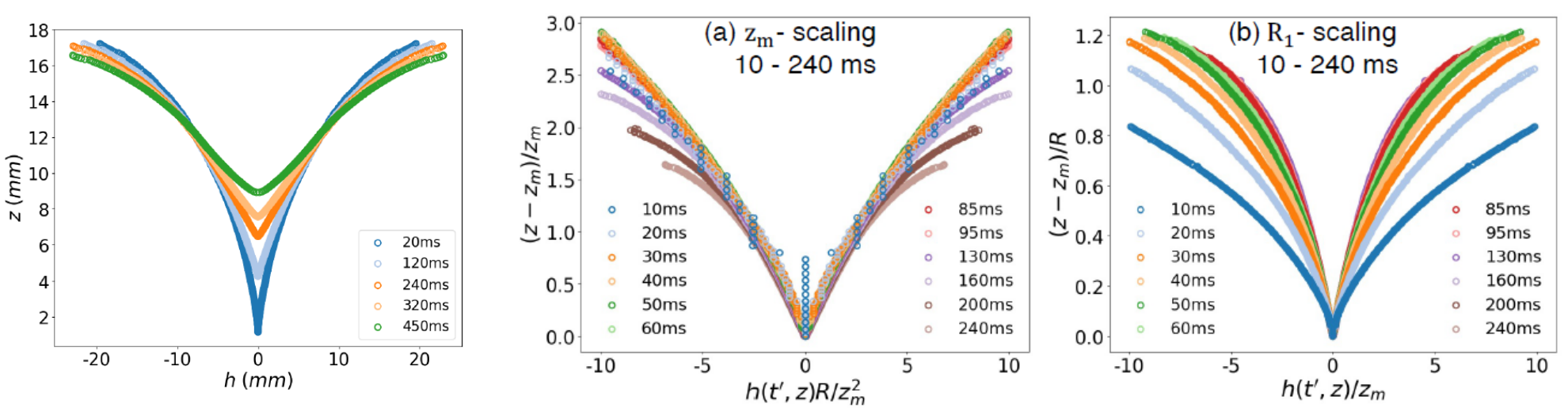}
\caption{Left: Representative interface profiles obtained for parameter set
1, $(e,D_{0},R,\protect\nu )=$ $(0.50$, $3.0$, $12.5$, $10)$. For clarity,
only selected profiles extracted from the 1000-fps image sequence are shown.
Right: Comparison of interface profile collapses based on different
scalings, using the same dataset as in Fig.~\protect\ref{Fig3}L obtained at $%
1000~\text{fps.}$ (a) Clear collapse near the tip by the $z_{m}$-scaling in
Eq.~(\protect\ref{eq1}). (b) Partial collapse by the $R_{1}$-scaling in Eq.~(%
\protect\ref{eq2}).}
\label{Fig3}
\end{figure*}

\subsection{Emergence of new self-similar structures}

We investigate the interface dynamics characterized by the shape function $%
h(t,z)$ defined in Fig.~\ref{Fig1}(b). Representative interface profiles
during recovery are shown in Fig.~\ref{Fig3}L for parameter set 1 in Table~%
\ref{tab:exp_conditions}. For visual clarity, only selected profiles
extracted from the 1-ms (1000 fps) image sequence are displayed. Throughout
this paper, the profiles are presented after left-right averaging.

Previous work by Yoshino and Okumura \cite{yoshino2025partial} demonstrated
that the near-tip dynamics obey the following self-similar form, which we
call the $z_{m}$-scaling: 
\begin{equation}
\frac{h(t^{\prime },z)}{z_{m}^{2}/R}=\Gamma \left( \frac{z-z_{m}}{z_{m}}%
\right) ,  \label{eq1}
\end{equation}%
with $\Gamma (x)\sim x$ for $x\ll 1$. This form describes the interface on
the scale $z-z_{m}\sim z_{m}$ in the vertical direction, as seen in the
argument of $\Gamma $ in Eq.~(\ref{eq1}), i.e., $\frac{z-z_{m}}{z_{m}}$.
Accordingly, we call the scaling in Eq.~(\ref{eq1}) "the $z_{m}$-scaling" in
the following.

While this scaling accurately captures the local quasi-linear structure near
the tip, the profiles in Fig.~\ref{Fig3}L indicate that the recovery process
extends over much larger scales, where non-linear structure is visible. We
therefore examine the larger scale $z-z_{m}\sim R$ for which Yoshino and
Okumura proposed the alternative self-similar form \cite{JPS2023autumn}, 
\begin{equation}
\frac{h(t^{\prime },z)}{z_{m}}=\Gamma \left( \frac{z-z_{m}}{R}\right) ,
\label{eq2}
\end{equation}%
with $\Gamma (x)\sim x$ for $x\ll 1$. Importantly, the two forms, Eqs.~(\ref%
{eq1}) and (\ref{eq2}) are mutually consistent because both reduce to $\frac{%
h(t^{\prime },z)}{z_{m}}\sim \frac{z-z_{m}}{R}$ in the vicinity of the tip.

We call the scaling form in Eq.~(\ref{eq2}) "the $R_{1}$-scaling" because
later we generally consider "the $R_{\beta }$-scaling" defined by a generic
form,%
\begin{equation}
\frac{h(t^{\prime },z)}{R(z_{m}/R)^{\beta }}=\Gamma \left( \frac{z-z_{m}}{R}%
\right) ,  \label{eqb}
\end{equation}%
which reduces to\ "the $R_{1}$-scaling" in Eq.~(\ref{eq2}) when $\beta =1$.
Motivated by the occurrence of simple fractional exponents in analytic
self-similar solutions of interfacial dynamics \cite{eggers2015singularities}%
, we explored simple rational values of $\beta $ and found distinct profile
collapses for $\beta =1/6$, 1/2, 1, and 2, as demonstrated below.

The two scaling descriptions, the $z_{m}$- and $R_{1}$-scalings, are
compared in Fig.~\ref{Fig3}R using the profiles shown in Fig.~\ref{Fig3}L.
As expected from the previous study, the $z_{m}$-scaling in Eq.~(\ref{eq1})
successfully captures the local dynamics near the tip, yielding a clear
collapse for $10\leq t^{\prime }\leq 240$ ms in the region $0<\frac{h}{%
z_{m}^{2}/R}\leq 4$, as seen in Fig.~\ref{Fig3}R(a), although the profiles
at $10~\text{ms }$appear to be affected by limited spatial resolution.

At larger scales, however, the $R_{1}$-scaling in Eq.~(\ref{eq2}) provides a
substantially broader spatial description of the interface. Although not
immediately apparent in Fig. \ref{Fig3}R(b), the collapse extends over a
substantially larger portion of the profile than that achieved by the $z_{m}$%
{}-scaling. This becomes evident when the earliest profiles ($t^{\prime
}<100~$ms) are excluded, as shown in Fig. \ref%
{fig:self-similarity_crossover1}(a), where an excellent collapse is obtained
for $100\leq t^{\prime }\leq 240$ ms in the region $0<\frac{h}{z_{m}}\leq 3$.

\begin{figure*}[!t]
\centering
\includegraphics[
width=129mm,
clip,
trim=180bp 205bp 178bp 205bp
]{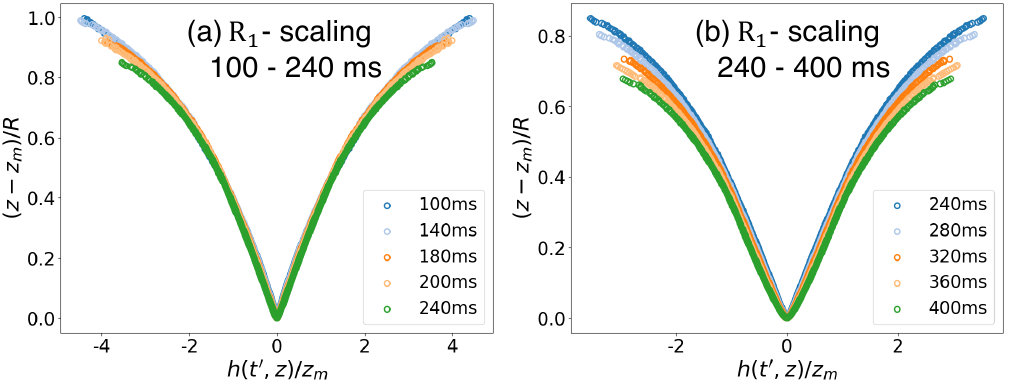}
\caption{Test of the $R_{1}$-scaling in \eqref{eq2}, using the same dataset
as in Fig.~\protect\ref{Fig3}L obtained at $1000~\text{fps}$: (a) Clear
collapse at earlier times ($100\leq t^{\prime }\leq 240~\text{ms}$). (b)
Deteriorated collapse at later times ($240\leq t^{\prime }\leq 400~\text{ms}$%
).}
\label{fig:self-similarity_crossover1}
\end{figure*}

These results demonstrate that extending the analysis from the local scale $%
z-z_{m}\sim z_{m}$ to the larger scale $z-z_{m}\sim R$ reveals an additional
self-similar structure governing a much broader portion of the recovering
interface while the temporal window becomes narrower. In fact, as shown in
Fig.~\ref{fig:self-similarity_crossover1}(b), the quality of the collapse
based on the $R_{1}$-scaling in \eqref{eq2} significantly deteriorates at $%
t^{\prime }\geq 240~\text{ms}$.

The breakdown of the $R_{1}$-scaling at late times signals a crossover to
another scaling regime, which we call the $R_{2}$-scaling following the
general definition in Eq.~(\ref{eqb}): 
\begin{equation}
\frac{h(t^{\prime },z)}{z_{m}^{2}/R}=\Gamma \left( \frac{z-z_{m}}{R}\right)
\label{eq:selfsimilar_afterbreakup_sheet_new2}
\end{equation}%
The validity of this scaling form is tested in Fig.~\ref%
{fig:self-similarity_crossover2}. While Fig.~\ref%
{fig:self-similarity_crossover2}(a) clearly shows that this new scaling form
does not work at earlier times ($t^{\prime }\leq 240\text{~ms}$), Fig.~\ref%
{fig:self-similarity_crossover2}(b) demonstrates that it provides an
excellent collapse for the profiles in the long-time region ($t^{\prime
}\geq 300\text{~ms}$). 
\begin{figure*}[!t]
\centering
\includegraphics[
width=138mm,
clip,
trim=160bp 200bp 160bp 195bp
]{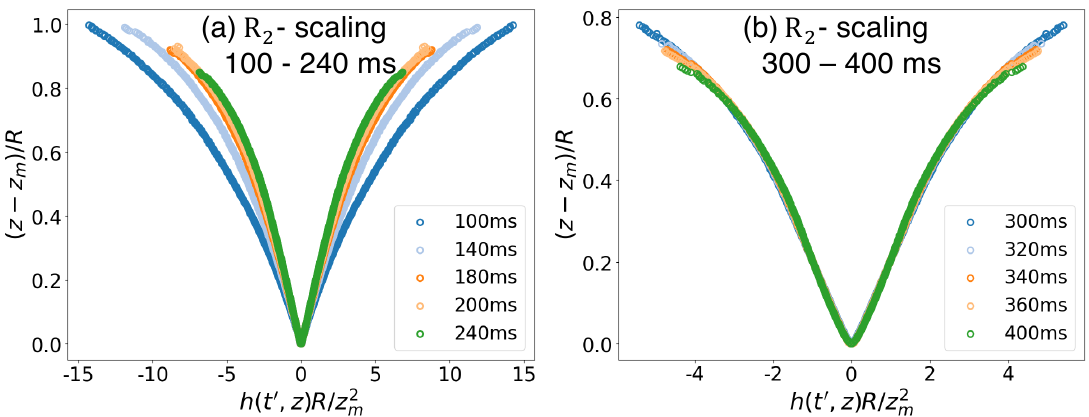}
\caption{Test of the $R_{2}$-scaling in 
\eqref{eq:selfsimilar_afterbreakup_sheet_new2} proposed for the long-time
region, using the same dataset as in Fig.~\protect\ref{Fig3}L obtained at $%
1000~\text{fps}$: (a) Limited collapse at earlier times ($100\leq t^{\prime
}\leq 240~\text{ms}$). (b) Clear collapse at later times ($300\leq t^{\prime
}\leq 400~\text{ms}$).}
\label{fig:self-similarity_crossover2}
\end{figure*}

We thus conclude that the present time-dependent interfacial profiles
exhibit a distinct scaling crossover from the $R_{1}$- to $R_{2}$-scaling,
i.e., Eq.~(\ref{eqb}) with $\beta =1$ to 2, as the system evolves far from
the breakup time at the length scale $z-z_{m}\sim R$.

\subsection{Multiple Crossovers\label{Hierarchical_Crossover}}

\begin{figure*}[!t]
\centering%
\includegraphics[
width=130mm,
clip,
trim=175bp 113.5bp 125bp 110bp
]{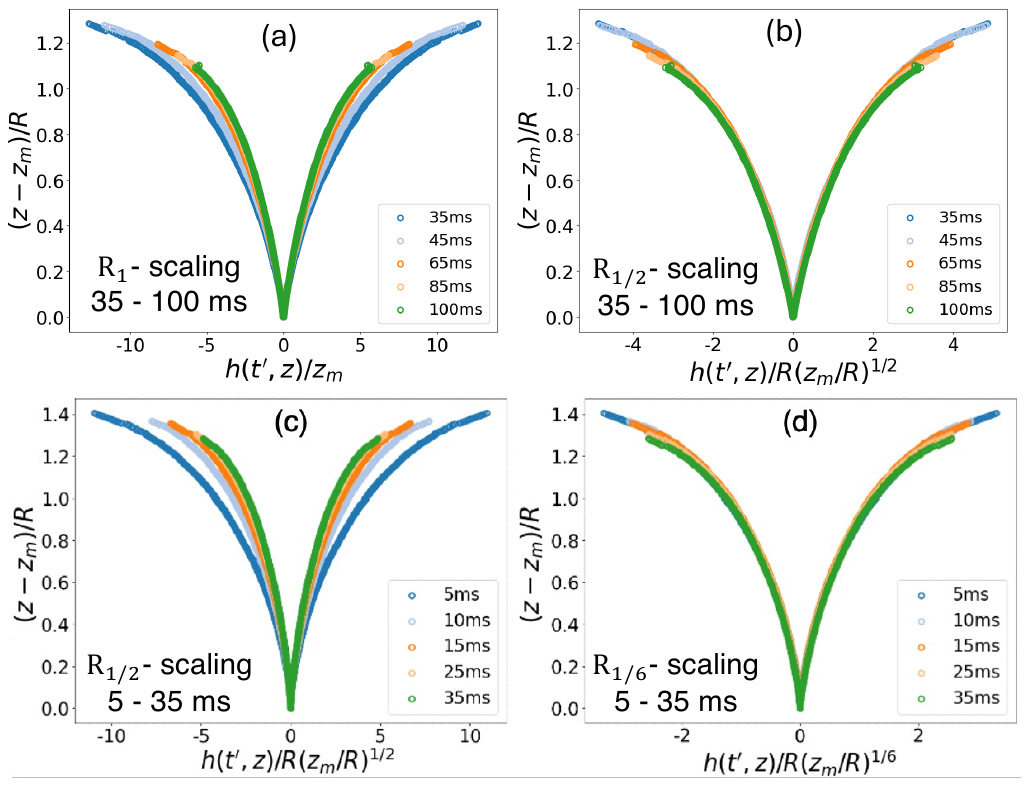}
\caption{Emergence of multiple self-similar crossovers in the early recovery
stage, demonstrated for the representative parameter set $(e,D_{0},R,\protect%
\nu )=(0.50,3.0,12.5,10)$. In the early stage ($35\,\text{ms}\leq t^{\prime
}\leq 100\,\text{ms}$), the $R_{1}$-scaling fails to produce a satisfactory
collapse (a), whereas the $R_{1/2}$-scaling yields a clear collapse (b). In
the ultra-early stage ($5\,\text{ms}\leq t^{\prime }\leq 35\,\text{ms}$),
the $R_{1/2}$-scaling breaks down (c), whereas the $R_{1/6}$-scaling
successfully collapses the profiles onto a common master curve (d).}
\label{fig:short_time_breakdown}
\end{figure*}

\begin{figure*}[!t]
\centering\includegraphics[width=0.8\textwidth]{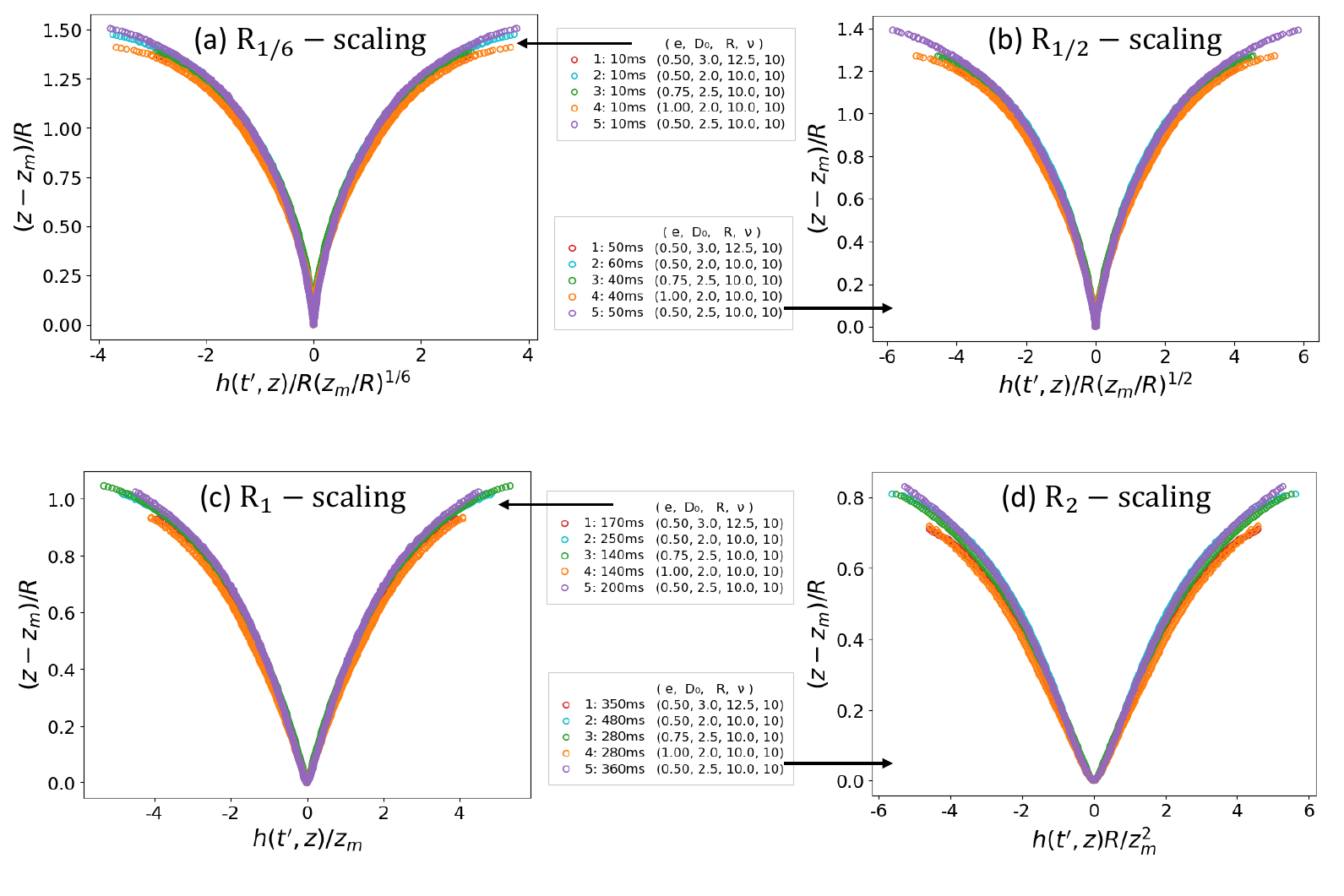}
\caption{Universality across the four scaling regimes. Interface-profile
collapses across variations in $e$, $D_{0}$, and $R$ for (a) the $R_{1/6}$%
{}-scaling, (b) the $R_{1/2}${}-scaling, (c) the $R_{1}${}-scaling, and (d)
the $R_{2}${}-scaling. Robust collapse persists near the tip in all four
regimes, while geometrical dependence becomes increasingly discernible away
from the tip.}
\label{fig:hierarchical_crossover_behavior}
\end{figure*}

The transition between the $R_{1}$- and $R_{2}$-scalings is only one of
multiple crossovers embedded in the dynamics. The deviations already visible
at early times ($t^{\prime }<100\,\mathrm{ms}$) in Fig.~\ref{Fig3}R(b) can
also be collapsed by considering two additional scalings, $R_{1/6}$- and $%
R_{1/2}$-scalings. Altogether, the dynamics exhibits three scaling
crossovers: from the $R_{1/6}$- to $R_{1/2}$-scaling, from the $R_{1/2}$- to 
$R_{1}$-scaling, and from $R_{1}$- to $R_{2}$-scaling.

Figure \ref{fig:short_time_breakdown} examines in detail the early-time ($%
t^{\prime }<100\,\mathrm{ms}$) deviations from the $R_{1}$ master curve
visible in Fig.~\ref{Fig3}R(b) over two time intervals, 35-100 ms and 5-35
ms. In the later interval (35-100 ms), the profiles fail to collapse under
the $R_{1}$-scaling as confirmed in (a), whereas they collapse under the $%
R_{1/2}$-scaling as shown in (b). Together with the $R_{1}$-scaling observed
at later times in Fig.~\ref{fig:self-similarity_crossover1}(a), comparison
of Fig.~\ref{fig:short_time_breakdown} (a) and (b) establishes the crossover
from the $R_{1/2}$- to $R_{1}$-scaling. In the earlier interval (5-35 ms),
the $R_{1/2}$-scaling breaks down as shown in (c), whereas the profiles
exhibit an excellent collapse under the $R_{1/6}$-scaling as shown in (d).
Together with the $R_{1/2}$-scaling observed at later times in Fig.~\ref%
{fig:short_time_breakdown} (b), comparison of Fig.~\ref%
{fig:short_time_breakdown} (c) and (d) establishes the crossover from the $%
R_{1/6}$- to $R_{1/2}$-scaling. These results demonstrate that the apparent
deviations from the $R_{1}$ master curve in Fig.~\ref{Fig3}R(b) reflect
additional self-similar regimes at earlier times rather than a breakdown of
self-similarity.

\section{Discussion}

We established three scaling crossovers among four distinct self-similar
regimes for Data 1 in Table 1. Each regime is identified not merely by
successful collapse under one scaling, but also by the breakdown of the
neighboring scaling description over the same temporal range. Using the same
procedure, we confirmed the same crossover sequence for Data 2-5. For all
four scaling regimes, the profiles from different datasets collapse well in
the vicinity of the tip. However, the spatial extent of this common collapse
varies among the regimes. The $R_{1/6}$- and $R_{1/2}$-scalings exhibit
robust collapse over relatively broad regions, although a slight deviation
becomes discernible away from the tip for Data 4, corresponding to the
largest film thickness ($e=1.0$ mm), as seen in Fig.~\ref%
{fig:hierarchical_crossover_behavior}. For the $R_{1}${}-scaling, the common
collapse is restricted to a narrower region near the tip, with the deviation
of Data 4 becoming more apparent away from the tip. For the $R_{2}$%
{}-scaling, the common collapse remains restricted to the near-tip region,
while an additional deviation becomes discernible away from the tip for Data
1, which has the largest disk thickness ($D_{0}=3.0$ mm).

The systematic deviations observed across space and scaling regimes can be
understood from two complementary perspectives: scale separation and the
distance from the breakup singularity. From the viewpoint of scale
separation, the influence of system-specific geometrical scales is
suppressed near the singular region and becomes detectable as the relevant
interfacial scale grows. Notably, $e<D_{0}$ in the present system.
Accordingly, a weak deviation associated with the smaller scale $e$ is
already discernible away from the tip in the $R_{1/6}$- and $R_{1/2}$%
{}-scalings, whereas an additional deviation associated with the larger
scale $D_{0}${} becomes discernible only in the $R_{2}${}-scaling. This
ordering is consistent with progressively larger geometrical scales
influencing the dynamics as scale separation becomes weaker. At the same
time, the same observations show that universal collapse extends over
broader regions in the earlier regimes closer to the breakup singularity,
whereas geometrical dependence becomes more evident as the system evolves
toward the later $R_{1}${}- and $R_{2}${}-regimes. This trend reflects the
general tendency for universality to become more pronounced on approaching
critical points and singularities, where sensitivity to system-specific
details is reduced.

This scale-dependent emergence of geometrical sensitivity extends earlier
observations of confinement-induced self-similarity and partial memory in
the same experimental system \cite{nakazato2018self, yoshino2025partial}.
More generally, it is consistent with the view that universality can be
characterized by which system-specific information is lost or retained
asymptotically \cite{Okumura2026memory}.

A characteristic length scale associated with two $R_{\beta }$-scalings, say
the $R_{\beta _{1}}$- and $R_{\beta _{2}}$-scalings, can be estimated by
requiring their characteristic horizontal scales to become comparable: $%
R(z_{m}/R)^{\beta _{1}}\simeq R(z_{m}/R)^{\beta _{2}}$. For $\beta _{1}\neq
\beta _{2}$, this condition gives%
\begin{equation}
z_{m}^{\mathrm{x}}\simeq R.  \label{eq:crossover_condition}
\end{equation}%
Remarkably, the resulting characteristic value $z_{m}^{\mathrm{x}}$ is
independent of both $\beta _{1}$ and $\beta _{2}$. Substituting $z_{m}^{%
\mathrm{x}}\simeq R$ into the scaling law $z_{m}/D\sim (\Delta \rho
gRt^{\prime }/\eta )^{\alpha }$ gives the corresponding characteristic time
scale 
\begin{equation}
t_{\mathrm{x}}^{(\alpha )}=\frac{\eta }{\Delta \rho gR}(\frac{R}{D}%
)^{1/\alpha }  \label{eq:time_scale}
\end{equation}%
The calculated values are summarized in Appendix \ref{A-C}. For each of Data
1-5, we determined the beginning and end of each scaling regime using the
same collapse-and-breakdown procedure as in Figs. 4-7 and compared the
resulting time windows after normalization by $t_{\mathrm{x}}^{(\alpha )}$%
{}. Guided by the crossover in the $z_{m}${} dynamics from $\alpha \simeq
4/5 $ to 1/2, which occurs near the onset of the $R_{1}${}-scaling, we use $%
t_{\mathrm{x}}^{(4/5)}${} to normalize the $R_{1/6}$- and $R_{1/2}$%
{}-regimes and $t_{\mathrm{x}}^{(1/2)}${} for the $R_{1}$- and $R_{2}$%
-regimes. The normalized time windows are consistently centered at
approximately $t=3t_{\mathrm{x}}^{(4/5)}$, $12t_{\mathrm{x}}^{(4/5)}$, $15t_{%
\mathrm{x}}^{(1/2)}$, and 30$t_{\mathrm{x}}^{(1/2)}${}, respectively.
Despite the limited range of $t_{\mathrm{x}}^{(\alpha )}${} accessible under
the present experimental conditions, the consistency of the normalized time
windows across Data 1-5 supports the interpretation of $t_{\mathrm{x}%
}^{(\alpha )}${}, derived from the geometric condition $z_{m}^{\mathrm{x}%
}\simeq R$, as a characteristic time scale for organizing the occurrence of
the different self-similar regimes.

The four self-similar regimes are characterized by the following sequence of
horizontal scales:

\begin{equation}
R\left( \frac{z_{m}}{R}\right) ^{\frac{1}{6}}\rightarrow R\left( \frac{z_{m}%
}{R}\right) ^{\frac{1}{2}}\rightarrow R\left( \frac{z_{m}}{R}\right)
\rightarrow R\left( \frac{z_{m}}{R}\right) ^{2}.  \label{eq:star_crossover}
\end{equation}%
It should be emphasized that this sequence does not imply a temporal
decrease of the physical horizontal length scale. Although the exponent $%
\beta $ increases from 1/6 to 2 while $z_{m}/R\ll 1$, the quantity $z_{m}/R$
simultaneously increases during recovery. Consistent with this, the
experimentally observed blunting of the interface indicates that the
effective horizontal scale increases with time, since the interface
sharpness scales as $(z-z_{m})/h\sim R/h$. An intriguing feature of %
\eqref{eq:star_crossover} is that the sequence can also be written as $%
Ra\rightarrow Ra^{3}\rightarrow Ra^{6}\rightarrow Ra^{12}$ with $%
a=(z_{m}/R)^{1/6}$. Whether this algebraic structure reflects a deeper
organizing principle underlying the multiple self-similar scalings remains
an open question.

More broadly, recent renormalization-group (RG) approaches to nonlinear PDEs
identify self-similar states with RG fixed points and relate universality to
the elimination of irrelevant structures \cite{bricmont1994renormalization,
Goldenfeld, chen1995numerical, Okumura2025RG, okumura2026combined,
Okumura2026nonlinear}. From this perspective, the successive self-similar
regimes observed here raise the question of whether the multiple crossovers
reflect transitions among distinct RG fixed points or asymptotic states.

\section{Conclusion\label{conclusion and outlook}}

In the present study, we investigated the post-breakup recovery dynamics of
a quasi-two-dimensional fluid interface in a Hele-Shaw system. By extending
the accessible temporal and spatial ranges beyond those examined previously,
we identified multiple self-similar crossovers during interface recovery.
Rather than approaching a single asymptotic self-similar state through a
single transition, the interface evolves successively through four distinct
scaling regimes, characterized by the $R_{1/6}$-, $R_{1/2}$-, $R_{1}$-, and $%
R_{2}$-scalings. The same crossover sequence persists across variations in
experimental parameters. Universal behavior remains robust near the
interface tip across all four regimes and extends over broader spatial
regions in the regimes closer to the breakup singularity.

The physical origin of the observed sequence, including the emergence of the
exponents 1/6, 1/2, 1, and 2, remains an open question. An intriguing
possibility is that the observed sequence reflects dynamical selection among
multiple self-similar solutions, analogous to selection mechanisms discussed
in singular free-surface flows \cite{eggers2015singularities,
1993PRLEggersPinchoff, 1994ScienceNagelDropFallingFaucet}. Whether
additional self-similar regimes exist, and what determines the dynamical
selection among them, remain important questions for future work.

\textit{The authors are grateful to Ikumi Yoshino for her contributions at
the early stage of this study. This work was supported by JSPS\ KAKENHI
Grant Number JP24K00596. }

\textit{AI Disclosure}: Microsoft 365 Copilot (GPT-5) was used to assist
with English-language editing and manuscript presentation. All AI-assisted
suggestions were critically reviewed and verified by the authors. No
generative AI was used to generate or analyze research data, figures, or
scientific results.

\appendix

\section{Experimental details: reproducibility and release system\label{App1}%
}

To ensure reproducible initial conditions, the disk was pre-wetted with the
same PDMS used to fill the cell before release. After each experimental run,
the disk was removed from the cell and all adhering PDMS was carefully
removed before the next trial. The liquid level was kept constant throughout
the experiments, and the disk was positioned so that approximately one-third
of its diameter was submerged prior to release.

To guide the disk during release, acrylic spacers with a length of
approximately 2 cm, a width of 1 cm, and a thickness $e$ were attached near
the opening of the cell. These spacers ensured that the disk was released
from the center of the cell and moved vertically through the middle of the
gap.

Compared with the previous study, the release mechanism was significantly
improved by introducing an automated electromagnetic system. Because the
disk was magnetic, an electromagnet (TMEH-A4, TRUSCO Nakayama Corp.)
controlled by a DC power supply (AD-8723D, A\&D Co., Ltd.) was installed
beside the cell (Fig.~\ref{Fig1}(a)). The disk was held by the electromagnet
at voltages between 16 and 24 V and released by switching off the power
supply. This procedure eliminated the variability associated with manual
release and substantially improved experimental reproducibility. The release
process may generate transient electromagnetic disturbances. Their possible
influence was examined experimentally, and no measurable effect on the
post-breakup dynamics was detected.

The images were captured by a high-speed camera (FASTCAM SA-X, Photron)
equipped with a macro lens (Micro NIKKOR 60-mm f/2.8G ED, Nikon) at a
resolution of 1280$\times $1024 pixels. A calibration factor was determined
separately for each experimental condition using ImageJ, with a typical
value ranging from $20$ to $25$~pixels/mm.

Although the interface appears as a band with finite thickness in the
images, all quantitative analyses presented in this study were performed
using the inner interface profile. The inner contour was extracted from the
recorded images and converted into quantitative data for subsequent
analysis. The pinch-off moment ($t^{\prime }=0\,\text{ms}$) was
operationally defined as the frame immediately preceding the one in which
the minimum width of the outer interface profile first reached zero. This
definition introduces a temporal uncertainty of less than 1 ms in the
determination of the actual pinch-off time $t_{c}$.

\section{Characteristic time scales\label{A-C}}

The time scales $t_{x}^{(1/2)}$ and $t_{x}^{(4/5)}$ calculated from Eq.~(\ref%
{eq:time_scale}) using $\alpha =1/2$ and 4/5, respectively, are summarized
in Table \ref{tabA}.

\begin{table}[tbh]
\caption{Characteristic time scales $t_{\mathrm{x}}^{(1/2)}$ and $t_{\mathrm{%
x}}^{(4/5)}$ for parameter sets in Table~\protect\ref{tab:exp_conditions}.}
\label{tabA}\centering%
\begin{tabular}{ccc}
\hline\hline
Index & $t_{x}^{(1/2)}~(\text{ms})$ & $t_{x}^{(4/5)}$ ($\text{ms}$) \\ \hline
1 & 11.5 & 4.88 \\ 
2 & 16.3 & 6.62 \\ 
3 & 9.2 & 4.62 \\ 
4 & 9.2 & 4.62 \\ 
5 & 12.0 & 5.46 \\ \hline\hline
\end{tabular}%
\end{table}


\end{document}